%% file: uchoo.tex
\documentclass[a4paper,10pt]{article}
\usepackage{graphicx}
\begin{document}

\title{\bf  Bounded Choice Queries for Logic Programming}
\author{Keehang Kwon\\
\sl \small Faculty of Computer Engineering, DongA  University\\
\sl \small 840 hadan saha, Busan, Korea\\
\small khkwon@dau.ac.kr
 }
\date{}
\maketitle

\input macros

\newcommand{\muprolog}{{Prolog$^{I}$}}
\newcommand{\uch}{uchoose}
\renewcommand{\pr}{ex}
\renewcommand{\prove}{ex} 


\noindent {\bf Abstract}: 
Adding versatile interactions to goals and queries in  logic programming is an 
essential task. 
Unfortunately, existing  logic languages can take input from the user only
via the $read$ construct. 
   We propose to add a new interactive goal  to allow for more controlled and more 
guided participation from the user. We illustrate our idea
via \muprolog, an extension of Prolog with bounded choice goals.

{\bf keywords:} interactive queries, logic programming, bounded choice, read.



\section{Introduction}\label{sec:intro}

Adding interactions to queries in  logic programming is  an essential task. 
The interactive queries that has been used
in logic programming has been restricted to the $read$ statement. 
The $read$ goal is  of the form $read(X) G$ where $G$ is a goal and 
$X$ can have any value. Hence, it is a form of an $unguided$ interactive
goal.  However, in several situations,
the system requires the user to choose  one among many alternatives. In other
words, they require a form of $guided$ and 
{\it bounded} interactions. Examples include most interactive systems.

The use of bounded interactions thus provides the user with a  facility in 
simulating interactive
systems. For this, this paper introduces a new goal statement $\uch(G_1,\ldots,G_n)$ 
where each  $G_i$ is a goal. This has the following execution semantics:

\[ \pr(D, \uch(G_1,\ldots,G_n))\ if\ \pr(D,  G_i) \] 

\noindent where $i$ is chosen by the user and $D$ is a program.
In the above definition, the system  requests the user to choose $i$ and then proceeds
with solving $G_i$.
It can be easily seen that our new statement has many applications
 in most interactive systems.

As an  example, let us consider a fast-food restaurant where you can have the
hamburger set or the fishburger set. For a hamburger set, you can have a hamburger and
a side-dish vegetable (onion or corn).
 For a fishburger set, you can 
have a fishburger and a side-dish vegetable (onion or corn). 
The menu is provided by the following definition:

\begin{exmple}
$price(h,3).$ \% hamburger, three dollars\\
$price(f,4).$ \% fishburger, four dollars \\
$price(o,1)$ \% onion, one dollar\\
$price(c,2)$ \% corn, two dollars \\
\end{exmple}
\noindent 
 As a particular example, consider a goal task 

   \begin{exmple}
$read(X) $ \\
 $read(Y)$  \\
$price(X,W) \land price(Y,Z)$\\
\end{exmple}
\noindent 
In our context,
  execution proceeds as follows: the system 
 requests the user to type in a particular burger and a vegetable. 
After they are selected, the system  computes their  prices ($W,Z$).
 Note that this goal is difficult to use and error-prone 
 because the user may type in an invalid value.

Our $\uch$ statement is useful to avoid this kind of human errors.
Consider a goal 

\begin{exmple}
$\uch ($ \\
\> $price(h,W) \land price(o,Z),$  \\
\> $price(f,W) \land price(o,Z),$\\
\> $price(h,W) \land price(c,Z),$  \\
\> $price(f,W) \land price(c,Z)$\\
$)$.
\end{exmple}
\noindent This goal expresses the task of the user choosing one among
four combinations.
 Note that this goal is much easier to use.
 The system now requests the user to select one (by typing 1,2,3 or 4)
among four -- rather than type in --
 combinations. After it is selected, 
the system produces their prices.

   To present our idea as simple as possible, this paper focuses on Prolog.
 In this paper we present the syntax and semantics of this extended language, 
show some examples of its use. 
The remainder of this paper is structured as follows. We describe 
our language
 in Section 2. In Section \ref{sec:modules}, we
present some examples of  \muprolog.
Section~\ref{sec:conc} concludes the paper.

\section{ Prolog with Bounded Choice Queries}\label{sec:logic1}

Our logic, an extended Horn clause logic, is described
by $G$- and $D$-formulas given by the syntax rules below:
\begin{exmple}
\>$G ::=$ \>  $A \sep   G \land  G \sep    \some x\ G \sep read(x) G  \sep  
\uch(G_1,\ldots,G_n)$   \\
\>$D ::=$ \>  $A  \sep G \supset A\ \sep \all x\ D \sep D \land D$\\
\end{exmple}
\noindent
In the rules above, $A$  represents an atomic formula.
A $D$-formula  is called a  Horn
 clause, or simply a clause. 

In the transition system below, $G$-formulas will function as 
queries and a  $D$-formula will constitute  a program. 
 We will  present the operational 
semantics for this language  as inference rules \cite{Khan87}, using the one
in \cite{MN12}. 
The rules for executing queries in our language are based on
uniform provability \cite{HM94,MNPS91}.  Note that execution  alternates between 
two phases: the goal reduction phase 
and the backchaining phase.

\begin{defn}\label{def:semantics}
Let $G$ be a goal and let $D$ be a program.  
Then   executing $G$ from $D$ -- written as $\prove(D,G)$ --
 is defined as follows: 

\begin{numberedlist}

\item  $bc(A,D,A)$. \%  a success.

\item    $bc(G\supset A,D,A)$ if 
 $\prove(D, G)$.

\item    $bc(\all x D_0,D,A)$ if   $bc([t/x]D_0,
D, A)$.

\item    $bc( D_0 \land D_1,D,A)$ if   $bc(D_0, D, A)$.

\item    $bc( D_0 \land D_1,D,A)$ if   $bc(D_1, D, A)$.

\item    $\prove(D,A)$ if   $bc(D,D, A)$.

\item $\prove(D,G_0 \land G_1)$ if $\prove(D,G_0)$  and 
  $\prove(D,G_1)$. 

\item $\prove(D,\exists x G)$  if  $\prove(D,[t/x]G)$. Typically selecting $t$ 
                      can be achieved via the unification process.

\item $\prove(D, read(x) G)$ if  $\prove(D, [kbd/x]G)$ where  $kbd$ is the
keyboard input.

\item $\prove(D, \uch(G_1,\ldots,G_n))$ if  $\prove(D, G_i)$  where  $i$  is chosen by the user.

\end{numberedlist}
\end{defn}

\noindent  
In the above rules, the symbol $\uch$  provides 
choice operations by the user.

\section{Examples }\label{sec:modules}

As an  example, let us consider the following database which contains the today's flight
information for major airlines such as Panam and Delta airlines.

\begin{exmple}
\% panam(source, destination, dp\_time, ar\_time) \\
\% delta(source, destination, dp\_time, ar\_time) \\
$panam(paris, nice, 9:00, 10:50)$\\
$\vdots$\\
$panam(nice, kiev, 9:45, 10:10)$\\
$\vdots$\\
$delta(paris, nice, 8:40, 09:35)$\\
$\vdots$\\
$delta(paris, kiev, 9:24, 09:50)$\\
\end{exmple}
\noindent Consider a (second-order) goal

\[ read(X)\  X(paris,nice,Dt,At). \]

\noindent This goal expresses the task of diagnosing whether the user has a 
flight in the airline X with their departure/arrival times to fly from Paris to Nice today.
The system then requests the user to type in a particular airline.
 Note that this goal is difficult to use and error-prone 
 because the user may type in an invalid airline.

Our $\uch$ statement is useful to avoid this kind of human errors.
Consider a goal 

\begin{exmple}
$\uch ($ \\
\> $ panam(paris,nice,Dt,At),$  \\
\> $ delta(paris,nice,Dt,At)$\\
$)$.
\end{exmple}
\noindent
This goal expresses the task of the user choosing one between Panam and Delta
 to fly from Paris to Nice today.
 Note that this goal is much easier to use than the above.
 The system now requests the user to select one among two -- rather than type in --
 airlines. After it is selected, 
the system produces the departure and arrival
time of the flight of the chosen airline.

\section{Conclusion}\label{sec:conc}

In this paper, we have considered an extension to Prolog with bounded choice queries.
 This extension allows goals of 
the form  $\uch(G_1,\ldots,G_n)$  where each $G_i$ is a goal.
This goal makes it possible for  Prolog
to model decision steps from the user.

We plan to connect our execution model to Japaridze's Computability Logic \cite{Jap03,Jap08}
which has many interesting applications (for example, see \cite{KHP13})
 in information technology.

\section{Acknowledgements}

This work  was supported by Dong-A University Research Fund.

\bibliographystyle{ieicetr}



\end{document}

%% file: macros.tex
\newenvironment{describe}{\begin{list}{}{\setlength\leftmargin{80pt}}\setlength\labelsep{10pt}\setlength\labelwidth{70pt}}{\end{list}}

\newenvironment{flag}{\begin{list}{\makebox[20pt]{\hss$\circ$\enspace}}
                                  {\labelwidth 20pt}}{\end{list}}



\newenvironment{numberedlist}
{\begin{list}{\makebox[20pt]{\hss(\arabic{itemno})\enspace}}
             {\usecounter{itemno}\labelwidth 20pt}}{\end{list}}

\newenvironment{alphabetlist}
{\begin{list}{\makebox[20pt]{\hss(\alph{itemno1})\enspace}}
             {\usecounter{itemno1}\labelwidth 20pt}}{\end{list}}

\newenvironment{romanlist}
{\begin{list}{\makebox[20pt]{\hss(\roman{itemno2})\enspace}}
             {\usecounter{itemno2}\labelwidth 20pt}}{\end{list}}

\newcounter{itemno}

\newcounter{itemno1}

\newcounter{itemno2}
\newcounter{lemma}
\newcounter{exno}

\newcounter{defno}







\newenvironment{defn}{\refstepcounter{defno}\medskip \noindent {\bf
Definition \thedefno.\ }}{\medskip}

\newenvironment{ex}{\refstepcounter{exno}\medskip \noindent {\bf
Example \theexno.\ }}{\medskip}

\newenvironment{millerexample}{
 \begingroup \begin{tabbing} \hspace{2em}\= \hspace{5em}\= \hspace{5em}\=
\hspace{5em}\= \kill}{
 \end{tabbing}\endgroup}

\newenvironment{wideexample}{
 \begingroup \begin{tabbing} \hspace{2em}\= \hspace{10em}\= \hspace{10em}\=
\hspace{10em}\= \kill}{
 \end{tabbing}\endgroup}

\newcommand{\sep}{\;\vert\;}

\newcommand{\ra}{\rightarrow}
\newcommand{\app}{\ }
\newcommand{\appt}{\ }
\newcommand{\tup}[1]{\langle\nobreak#1\nobreak\rangle}

\newcommand{\hu}{{\cal H}^+}
\newcommand{\Free}{{\cal F}}
\newcommand{\oprove}{\vdash\kern-.6em\lower.7ex\hbox{$\scriptstyle O$}\,}
\newcommand{\true}{\top}

\newcommand{\Dscr}{{\cal D}}
\newcommand{\Pscr}{{\cal P}}
\newcommand{\Gscr}{{\cal G}}
\newcommand{\Fscr}{{\cal F}}
\newcommand{\Vscr}{{\cal V}}
\newcommand{\Uscr}{{\cal U}}
\newcommand{\pderivation}{{\cal P}\kern -.1em\hbox{\rm -derivation}}
\newcommand{\pderivationl}{{\cal P}\kern -.1em\hbox{\em -derivation}}
\newcommand{\pderivable}{{\cal P}\kern -.1em\hbox{\rm -derivable}}
\newcommand{\pderivablel}{{\cal P}\kern -.1em\hbox{\em -derivable}}
\newcommand{\pderivations}{{\cal P}\kern -.1em\hbox{\rm -derivations}}
\newcommand{\pderivability}{{\cal P}\kern -.1em\hbox{\rm -derivability}}
\newcommand{\eqm}[1]{=_{\scriptscriptstyle #1}}
\newcommand\subsl{\preceq}
\newcommand{\fnrestr}{\uparrow}

\newcommand{\match}{{\rm MATCH}}
\newcommand{\triv}{{\rm TRIV}}
\newcommand{\imit}{{\rm IMIT}}
\newcommand{\proj}{{\rm PROJ}}
\newcommand{\simpl}{{\rm SIMPL}}
\newcommand{\failed}{{\bf F}}

\newcommand{\Dsiginst}[1]{{[#1]_\Sigma}}
\newcommand{\Psiginst}[1]{{[#1]_\Sigma}}
\newcommand{\lnorm}{{\lambda}norm}
\newcommand{\seq}[2]{#1 \supset #2}
\newcommand{\dseq}[2]{#1_1,\ldots,#1_{#2}}

\newcommand{\all}{\forall}
\newcommand{\some}{\exists}
\newcommand{\lambdax}[1]{\lambda #1\,}
\newcommand{\somex}[1]{\some#1\,}
\newcommand\allx[1]{\all#1\,}

\newcommand{\subs}[3]{[#1/#2]#3}
\newcommand{\rep}[3]{S^{#2}_{#1}{#3}}
\newcommand{\ie}{{\em i.e.}}
\newcommand{\eg}{{\em e.g.}}

\newcommand{\lbotr}{$\bot$-R}
\newcommand{\ldbotr}{\bot\mbox{\rm -R}}
\newcommand{\landl}{$\land$-L}
\newcommand{\ldandl}{\land\mbox{\rm -L}}
\newcommand{\landr}{$\land$-R}
\newcommand{\ldandr}{\land\mbox{\rm -R}}
\newcommand{\lorl}{$\lor$-L}
\newcommand{\ldorl}{\lor\mbox{\rm -L}}
\newcommand{\lorr}{$\lor$-R}
\newcommand{\ldorr}{\lor\mbox{\rm -R}}
\newcommand{\limpl}{$\supset$-L}
\newcommand{\ldimpl}{\supset\mbox{\rm -L}}
\newcommand{\limpr}{$\supset$-R}
\newcommand{\ldimpr}{\supset\mbox{\rm -R}}
\newcommand{\lnegl}{$\neg$-L}
\newcommand{\ldnegl}{\neg\mbox{\rm -L}}
\newcommand{\ldnegr}{\neg\mbox{\rm -R}}
\newcommand{\lalll}{$\forall$-L}
\newcommand{\ldalll}{\forall\mbox{\rm -L}}
\newcommand{\lallr}{$\forall$-R}
\newcommand{\ldallr}{\forall\mbox{\rm -R}}
\newcommand{\lsomel}{$\exists$-L}
\newcommand{\ldsomel}{\exists\mbox{\rm -L}}
\newcommand{\lsomer}{$\exists$-R}
\newcommand{\ldsomer}{\exists\mbox{\rm -R}}
\newcommand{\ldlamlr}{\lambda}
\newcommand{\sequent}[2]{\hbox{{$#1\ \longrightarrow\ #2$}}}
\newcommand{\prog}[2]{\hbox{{$#1\ \supset\ #2$}}}
\newcommand{\run}{\Gamma}

\newcommand{\Ibf}{{\bf I}}
\newcommand{\Cbf}{{\bf C}} 
\newcommand{\Cbfpr}{{\bf C'}}

\newcommand{\cprove}{\vdash_C}
\newcommand{\iprove}{\vdash_I}

\newsavebox{\lpartfig}
\newsavebox{\rpartfig}


\newenvironment{exmple}{
 \begingroup \begin{tabbing} \hspace{2em}\= \hspace{3em}\= \hspace{3em}\=
\hspace{3em}\= \hspace{3em}\= \hspace{3em}\= \kill}{
 \end{tabbing}\endgroup}
\newenvironment{example2}{
 \begingroup \begin{tabbing} \hspace{8em}\= \hspace{2em}\= \hspace{2em}\=
\hspace{10em}\= \hspace{2em}\= \hspace{2em}\= \hspace{2em}\= \kill}{
 \end{tabbing}\endgroup}

\newenvironment{example}{
\begingroup  \begin{tabbing} \hspace{2em}\= \hspace{3em}\= \hspace{3em}\=
\hspace{3em}\= \hspace{3em}\= \hspace{3em}\= \hspace{3em}\= \hspace{3em}\= 
\hspace{3em}\= \hspace{3em}\= \hspace{3em}\= \hspace{3em}\= \kill}{
 \end{tabbing} \endgroup }

\newcommand{\sand}{sand} 
\newcommand{\pand}{pand} 
\newcommand{\cor}{cor} 

\newcommand{\lb}{\langle}
\newcommand{\rb}{\rangle}
\newcommand{\pr}{prov}
\newcommand{\prG}{intp}
\newcommand{\prSG}{intp_E}
\newcommand{\intp}{intp_o}
\newcommand{\prove}{exec} 
\newcommand{\np}{invalid} 
\newcommand{\Ra}{\supset}  
\newcommand{\add}{\oplus} 
\newcommand{\adc}{\&} 
\newcommand{\Cscr}{{\cal C}}
\newcommand{\seqweb}{SProlog}
\newcommand{\sprog}{{SProlog}}

\newtheorem{theorem}[lemma]{Theorem}

\newtheorem{proposition}[lemma]{Proposition}

\newtheorem{corollary}[lemma]{Corollary}
\newenvironment{proof}
     {\begin{trivlist}\item[]{\it Proof. }}%
     {\\* \hspace*{\fill} \end{trivlist}}

\newcommand{\seqand}{\prec}
\newcommand{\seqor}{\cup}
\newcommand{\seqandq}[2]{\prec_{#1}^{#2}}
\newcommand{\parandq}[2]{\land_{#1}^{#2}}
\newcommand{\exq}[2]{\exists_{#1}^{#2}}
\newcommand{\ext}{intp_G}